\documentclass{iopart}

\usepackage{epsfig}

\newcommand{\sgn}{{\rm sgn}}
\newcommand{\bt}{\bar \theta}
\newcommand{\bsi}{{\bar \sigma}_+}
\newcommand{\bqz}{{\bar Q}_0}
\newcommand{\bqp}{{\bar Q}_+}
\newcommand{\bc}{{\bar C}_1}

\begin{document}

\jl{6}

\title{Spatially self-similar spherically symmetric perfect-fluid 
models}

\author{Martin Goliath \dag, Ulf S Nilsson \dag \ and 
        Claes Uggla \ddag \\
{\dag \ Department of Physics, Stockholm University, Box 6730,
  S-113 85 Stockholm, Sweden} \\
{\ddag \ Department of Physics, Lule{\aa} University of
  Technology, S-951 87 Lule{\aa}, Sweden}}
 
\begin{abstract}
Einstein's field equations for spatially self-similar spherically symmetric 
perfect-fluid models are investigated. The field equations are rewritten as a 
first-order system of autonomous differential equations. Dimensionless 
variables are chosen in such a way that the number of equations in the 
coupled system is reduced as far as possible and so that the reduced phase 
space becomes compact and regular. The system is subsequently analysed 
qualitatively with the theory of dynamical systems. 
\end{abstract}

\pacs{0420, 0420J, 0440N, 9530S, 9880H}

\maketitle

\section{Introduction}
Spherically symmetric self-similar perfect fluid spacetimes have attracted 
considerable attention during the last couple of decades (see e.g., 
\cite{caryah,carcoley} for references). This is not surprising since they 
lead to ordinary differential equations while still providing an arena for a 
number of interesting physical phenomena: shock waves 
\cite{cahtau71,bogoyavlensky}; self-similar perturbations of the flat 
Friedmann-Lemaitre-Robertson-Walker (FLRW) solution, which are of relevance 
for the growth of primordial black holes and evolution of voids (see, e.g., 
\cite{caryah,carr,bicknell}); and violation of cosmic censorship 
\cite{oripir}, are some examples. These models also have broader 
implications. For example, they have turned out to be of essential importance 
when it comes to understanding spherically symmetric black hole formation 
\cite{koike,maison}.

Self-similar spherically symmetric perfect-fluid models exhibit several 
preferred geometric structures. There are preferred directions associated 
with the 4-velocity of the fluid and the homothetic Killing vector (with 
self-similar spacetimes here we mean spacetimes with a homothetic Killing 
vector. For a discussion about self-similarity, see \cite{coley}). Another 
preferred structure is the area of the spherical symmetry surfaces. These 
three structures have led to three approaches, each using a specific 
coordinate gauge. The first is `the comoving perfect-fluid approach' 
initiated by Cahill and Taub \cite{cahtau71}. The second is `the homothetic 
approach' used by Bogoyavlensky and co-workers 
\cite{bogoyavlensky,moschetti,anile}. The third is `the Schwarzschild 
approach' used by, for example, Ori and Piran (see, e.g., \cite{oripir}). 
Each gauge suggests more or less natural variables and each gauge has its 
individual physical interpretational advantages. Thus they are all 
complementary. However, it is possible to express many features gauge 
invariantly. This makes it possible to algebraically relate the variables one 
has used in the various approaches (this will be done in \ref{sec:coord}). It 
is worth mentioning that apart from the above three approaches there are also 
some other approaches, useful for studying the spacetime structure `near' 
singularities and for investigating global spacetime features, based on 
synchronous coordinate systems \cite{bogoyavlensky} and null coordinates 
\cite{patel}.

We will use the homothetic approach as our starting point. The homothetic 
approach has the advantage that it reveals a considerable structural 
similarity between the field equations for the self-similar spherically 
symmetric models and the hypersurface homogeneous models (e.g. spatially 
homogeneous (SH) models). There is a considerable literature about how one can 
treat the field equations of the hypersurface homogeneous models. Thus the 
homothetic approach makes it possible to transfer ideas from the hypersurface 
homogeneous arena to the present situation. However, the homothetic approach, 
where one chooses a coordinate along the orbits of the homothetic Killing 
vector (see, e.g., \cite{ear,wu,hhss}), has a disadvantage. In general, the 
symmetry surface changes causality. If one uses a homothetic diagonal gauge 
for the spherically symmetric case, then the spacetime must be covered with 
several coordinate patches; one for the region where the homothetic Killing 
vector is spacelike and one where it is timelike. Then one has to join the 
regions where the homothetic Killing vector is null. 

This is the first paper in a series with the aim of obtaining a global 
picture of the space of solutions for self-similar spherically symmetric 
perfect-fluid models. Here we will consider the case where the homothetic 
Killing vector is spacelike; the so-called spatially self-similar (SSS) 
spherically symmetric models. These models are characterized by a 
four-dimensional homothetic symmetry group acting multiply transitively on 
three-dimensional spatial surfaces. In a subsequent paper \cite{art:TSS} we 
will deal with the timelike self-similar (TSS) spherically symmetric 
perfect-fluid models, i.e. the models where the homothetic Killing vector is 
timelike. Most of the physically interesting phenomena, e.g. shock waves, are 
actually associated with the TSS region, and consequently it is not 
surprising that most previous work has focused on this region. However, to 
obtain a global understanding one also needs to consider the spatial region. 
In a third paper we will address the specific problem of matching solutions 
between the two regions as well as focusing more on physical properties of the 
space of solutions. Therefore the mathematical nature of the present paper 
is justified as the results are needed in the subsequent studies. However, 
one physically interesting property pertaining to the SSS region, density 
perturbations of the flat FLRW solution, is treated in the present paper.

By using the homothetic approach and ideas from hypersurface homogeneous 
models we will reformulate Einstein's field equations so that the dynamics 
takes place in a reduced and compact phase space; one system for the spacelike 
case and one for the timelike case. Moreover, it will turn out that in our 
formulation, all equilibrium points (critical points, singular points, fixed 
points) are hyperbolic (here we also include the case of a line of 
equilibrium points where there is one zero eigenvalue corresponding to the 
line). This is in stark contrast to earlier treatments where noncompact 
variables have been used and where certain parts of phase space have been 
`crushed'. To deal with these problems, one has made numerous variable 
changes and special treatments of various parts of phase space (see, e.g. 
\cite{bogoyavlensky}). This has resulted in an incomplete, or even 
misleading, picture of the global solution structure.

The line element for SSS spherically symmetric models, written in diagonal 
form where one of the coordinates is adapted to the homothetic symmetry, 
takes the form \cite{bogoyavlensky}
\begin{equation}
   d{\tilde s}^2 = e^{2x} ds^2 = e^{2x}\left[
   -dt^2 + D_1^2(t) dx^2 + D_2^2(t)\left(d\theta^2 +
   \sin^2(\theta)d\varphi^2 \right)\right] .
\end{equation}

We will consider perfect fluid models. The energy momentum tensor,
${\tilde T}_{ab}$, is thus given by
\begin{equation}
  {\tilde T}_{ab} = {\tilde \mu}u_a u_b + {\tilde p}(u_a u_b + 
                    {\tilde g}_{ab}) ,
\end{equation}
where $\tilde{\mu}$ is the energy density, $\tilde{p}$ is the pressure and
$u^a$ the 4-velocity of the fluid. We will assume
\begin{equation}
   \tilde{p}=\left( \gamma-1 \right)\tilde{\mu} ,
\end{equation}
as an equation of state, where the parameter $\gamma$ takes values in the 
interval $1 < \gamma < 2$ which include radiation ($\gamma = \case43$). 
Consequently, we have excluded dust ($\gamma = 1$) and stiff fluids 
($\gamma = 2$). These models behave quite differently compared to those in 
the above interval, and thus need special treatment. The dust solutions are 
known explicitly (these models are just special cases of the general 
spherically symmetric dust solutions, which are all known explicitly, see 
e.g., \cite{kramer}).

The outline of the paper is the following. In section 2 Einstein's field 
equations are rewritten in terms of a dimensionless set of variables in order 
to obtain a maximal reduction of the coupled system of ordinary differential 
equations. The variables are chosen so that they take values in a compact 
phase space. In section 3 the equations are subsequently analysed by means of 
methods from the theory of dynamical systems. In section 4 a numerical 
investigation is undertaken and global dynamical features are considered. We 
end with a discussion in section 5. Appendix A describes the properties of 
the fluid congruence, the condition for the spacetimes to belong to Petrov 
type 0 and also some expressions relevant for the mass function and density 
perturbations. Appendix B gives the relation between various coordinates and 
the relation between the present variables and other variables used in the 
literature.

\section{The dynamical system}
Following \cite{ssslrs} we introduce
\begin{eqnarray}
  D_1 &=& B_1{}^{-1} = e^{\beta^0 - 2\beta^+} , 
  \qquad D_2 = B_2{}^{-1} = e^{\beta^0 + \beta^+} , \\
  \theta &=& 3{\dot \beta}^0,\qquad \sigma_+ = 3{\dot \beta}^+ , \nonumber
\end{eqnarray}
where the dot denotes $d/dt$. The quantities $\theta$ and $\sigma_+$ describe 
the kinematical properties of the normal congruence of the symmetry surfaces 
in the $(M,ds^2)$ spacetime that is conformally related to the physical 
spacetime $(M,d{\tilde s}^2)$ with the homothetic factor (see, e.g., 
\cite{hhss}); $\theta$ is the expansion while $\sigma_+$ describes the shear. 
The orthonormal frame components, in $(M,ds^2)$, of the fluid velocity are 
conveniently parametrized by $(1,v,0,0)/\sqrt{1 - v^2}$, where $v$ is just 
the 3-velocity with respect to the symmetry surfaces. Einstein's field 
equations, ${\tilde G}_{ab} = {\tilde T}_{ab}$, and the conservation 
equations, ${\tilde T}^{ab}\!_{;b} = 0$, lead to a set of equations presented 
in \cite{ssslrs}. However, for mathematical reasons we will now go over to 
another set of variables. The motivation lies in a simplification of the 
constraint while keeping the `canonical quadratic' nature of the defining 
equation for $\mu_n$ (see the equations below). The new variables are given by
\begin{eqnarray}
  \bt &=& \frac{1}{\sqrt 3}(2\theta - \sigma_+) , \qquad
  \bsi = \frac{1}{\sqrt 3}(-\theta + 2\sigma_+) , \\
  \theta &=& \frac{1}{\sqrt 3}(2\bt + \bsi) , \qquad
  \sigma_+ = \frac{1}{\sqrt 3}(\bt + 2\bsi) . \nonumber
\end{eqnarray}
This leads to:

\paragraph{Evolution equations}
\begin{eqnarray}
  \dot{\bt} &=& -\frac{1}{\sqrt 3}\left[
  \bt^2 + \bsi^2 + \bt\bsi - 3B_1^2 +
  \frac{3(\gamma - 1)(1 - v^2)}
  {1+\left(\gamma-1\right)v^2}\mu_n\right] , \nonumber \\
  \dot\bsi &=& -\frac{1}{\sqrt 3}\left[
  \bsi^2 + 2\bt\bsi +
  \frac{3}{2}\frac{(2 - \gamma) + (3\gamma - 2)v^2}
  {1+\left(\gamma-1\right)v^2}\mu_n \right] ,
  \nonumber\\
  \dot{B_1} &=&\frac{1}{\sqrt 3} \bsi B_1 , \\
  \dot{B_2} &=&-\frac{1}{\sqrt 3}(\bt + \bsi)B_2 , \nonumber \\
  \dot{v} &=& \frac{1 - v^2}{{\sqrt 3}\gamma\left(1 - (\gamma - 1)v^2\right)}
  \left\{\gamma \left[ 2(\gamma - 1)\bt + \gamma\bsi \right] v \right.
  \nonumber \\
  & & \left.+ {\sqrt 3}\left[(\gamma - 1)(3\gamma - 2)v^2 -
  (2 - \gamma)\right]B_1 \right\} . \nonumber
\end{eqnarray}

\paragraph{Constraint equation}
\begin{equation}
   \gamma v\mu_n - \frac{2}{\sqrt 3}
   \left[ 1+\left(\gamma-1\right)v^2 \right]\bsi B_1 = 0 .
\end{equation}

\paragraph{Defining equation for $\mu_n$}
\begin{equation}\label{eq:muns2}
  \mu_n = \frac{1}{3}\left(\bt^2 -\bsi^2 - 3B_1^2 + 3B_2^2 \right) .
\end{equation}

\paragraph{Auxiliary equation}
\begin{equation}
  \dot{\mu}_n= \frac{-\gamma\mu_n}
  {\sqrt 3\left[1+\left(\gamma-1\right)v^2\right]}
  \left[ 2\bt + (1 - v^2)\bsi + 2{\sqrt 3}vB_1 \right] .
\end{equation}
The quantity $\mu_n$ is the energy density of the fluid measured
by an observer with the 4-velocity orthogonal to the symmetry surfaces
in $(M,ds^2)$, see \cite{hhss}.

As the final step we will introduce a new set of variables where the scale
invariance is used to decouple one equation. In spatially homogeneous 
cosmologies one often uses the expansion for this purpose. However, a very 
desirable property is to obtain a reduced phase space which is compact. 
Unfortunately, using, for example, $\theta$ or $\bt$ does not achieve this in 
the present case. Instead we will use the quantity $\bt^2 + 3B_2^2$ since 
this is a `dominant quantity' (see equation (\ref{eq:muns2}) and note that we 
assume a nonnegative energy density). We define a new variable $Y$,
\begin{equation}\label{eq:Y}
  Y = \sqrt{\bt^2 + 3B_2^2} , \qquad 
  B_2 = \sqrt{(Y^2 - \bt^2)/3} .
\end{equation}
Unfortunately, this will lead to a more mathematical treatment than if one
uses a more `physical' quantity instead of $Y$. However, there seems to be
no such quantity which leads to a compact phase space. Instead one is forced
to algebraically relate interesting physical objects, e.g. fluid kinematic 
quantities and mass functions to the variables used in this paper. These 
relations are given in \ref{sec:physquant}. We also introduce the 
$Y$-normalized bounded dimensionless variables
\begin{equation}
  \bqz = \frac{\bt}{Y} , \qquad  
  \bqp = \frac{\bsi}{Y} , \qquad
  \bc = \frac{{\sqrt 3}B_1}{Y} .
\end{equation}
Note that  $\bc$ is positive. The density $\mu_n$ is replaced by the density 
parameter $\Omega_Y$, which is defined by
\begin{equation}
  \Omega_Y=\frac{3\mu_n}{Y^2} .
\end{equation}
An introduction of a dimensionless time variable $\bar \tau$
\begin{equation}
  dt = {\sqrt 3}Y^{-1}d{\bar \tau} ,
\end{equation}
leads to a decoupling of the $Y$-equation
\begin{equation}
  Y^\prime = F_Y Y , \qquad
  F_Y = -\left[ \bqp + \bqz\left(2\bqp^2 + 
  \frac{\gamma}{1+\left(\gamma-1\right)v^2}\Omega_Y\right)\right] ,
\end{equation}
where a prime denotes $d/d{\bar \tau}$. The remaining coupled evolution 
equations can now be written in dimensionless form:

\paragraph{Evolution equations}
\begin{eqnarray}
  \bqz^{\prime} &=& -(1 - \bqz^2)\left[\bqp^2 - \bc^2 + 
  \frac{(\gamma - 1)(1 - v^2)}{1+\left(\gamma-1\right)v^2}\Omega_Y \right] ,
  \nonumber \\
  \bqp^{\prime} &=& -\bqz\bqp\left[2(1 - \bqp^2) - 
     \frac{\gamma\Omega_Y}{1+\left(\gamma-1\right)v^2}\right] \nonumber \\
  & & \qquad -\frac12 \frac{(2 - \gamma) + (3\gamma - 2)v^2}
  {1+\left(\gamma-1\right)v^2} \Omega_Y , \\
  \bc^{\prime} &=& 2\bc\left[\bqp + \bqz\bqp^2 +
  \frac{1}{2} \frac{\gamma}{1+\left(\gamma-1\right)v^2}\bqz\Omega_Y \right] ,
  \nonumber \\
  v^{\prime} &=& \frac{1-v^2}{\gamma\left[1 - (\gamma - 1)v^2\right]} 
  \left\{\gamma \left[ 2(\gamma - 1)\bqz + \gamma\bqp \right] v +\right. 
  \nonumber \\
  & & \left.\left[(\gamma - 1)(3\gamma - 2)v^2 - 
  (2 - \gamma)\right]\bc\right\} . \nonumber
\end{eqnarray}

\paragraph{Constraint equation}
\begin{equation}\label{eq:sssco}
  G \equiv \gamma v \Omega_Y - 2(1+\left(\gamma-1\right)v^2)\bqp\bc = 0 .
\end{equation}

\paragraph{Defining equation for $\Omega_Y$}
\begin{equation}\label{eq:omega}
  \Omega_Y = 1 - \bqp^2 - \bc^2 .
\end{equation}

\paragraph{Auxiliary equation}
\begin{equation}
  \Omega_Y^\prime = -\Omega_Y\left\{\frac{\gamma}{1+\left(\gamma-1\right)v^2}
  \left[2\bqz + (1 - v^2)\bqp + 2v\bc\right] + 2F_Y\right\} .
\end{equation}
The reduced equations and the equation for $Y$ are invariant under the 
transformation
\begin{equation}\label{eq:disc}
  ({\bar \tau},\bqz,\bqp,\bc,v) \rightarrow  
  (-{\bar \tau},-\bqz,-\bqp,\bc,-v) .
\end{equation}
The line element can be obtained when $\bqz,\bc$ and $Y$ have been found
through the relations
\begin{equation}
  D_1 = \sqrt{3}(Y \bc)^{-1} , \quad   
  D_2 = \sqrt{3}[Y^2(1 - \bqz^2)]^{-1/2} , \quad
  t = {\sqrt 3}\int \frac{d{\bar \tau}}{Y} .
\end{equation}
Because of the above discrete symmetry, there are, in general, {\it two}
orbits in the reduced phase space, related by (\ref{eq:disc}), corresponding 
to a single line element. There is also a class of orbits that are invariant 
under the above discrete symmetry and in this case each orbit corresponds to 
a single line element.

\section{Dynamical systems analysis of the reduced phase space}
The reduced phase space is determined by ($\bqz,\bqp,\bc,v$), related by the
constraint $G = 0$, given by equation (\ref{eq:sssco}). The boundary of the 
physical phase space is of essential importance in understanding the dynamics 
of the interior phase space and hence we will include it in our analysis and 
thus obtain a compactified phase space. We will not solve the constraint 
globally since this leads to problems. Instead we will follow 
\cite{ssslrs,hewitt} and solve it locally around the equilibrium points (i.e. 
we will solve the linearized constraint for different variables at different 
equilibrium points). This formulation will enable us to achieve a good 
understanding of the global structure of the reduced phase space.

\subsection{Equilibrium points and local analysis}
Here we will present the equilibrium points together with $\Omega_Y$, which 
will indicate if a point is a vacuum point or not, the gradient of the 
constraint and what variable we have locally solved for, and finally the 
eigenvalues. There are numerous equilibrium points. However, they are often 
closely related. We will use the following notation: the equilibrium points 
are denoted as 
\begin{equation}
  _{\sgn{({\bqz})}}{\rm Kernel}_{\sgn{({\bqp})}}^{\sgn (v)} .
\end{equation}
When there is no risk for confusion we will suppress $\sgn{({\bqz})}$, 
$\sgn{({\bqp})}$ or $\sgn (v)$.

\subsubsection{Equilibrium points with zero tilt, $v=0$.}

\paragraph{The equilibrium points $_\pm F$:}
\begin{eqnarray}
  \bqz&=&\pm1 , \quad 
  \bqp = -\frac{\bqz}{2} , \quad 
  \bc=0 ; \quad 
  \Omega_Y= \frac{3}{4} ; \nonumber \\
  \nabla G &=& (0,0,\bqz,\frac{3\gamma}{4}) , \quad 
  (\bc \ {\rm eliminated}) ; \\
  \lambda_1 &=& \frac{1}{2}(3\gamma-2)\bqz , \quad 
  \lambda_2 = \frac{-3}{4}(2-\gamma)\bqz , \quad
  \lambda_3 = \frac{1}{4}(3\gamma-2)\bqz . \nonumber
\end{eqnarray}
These points are saddle points in the full phase space but there is a 
two-dimensional separatrix surface, spanned by the eigenvectors corresponding 
to the eigenvalues $\lambda_1$ and $\lambda_3$ above, the orbits of which lie 
in the interior of phase space. 

\paragraph{The equilibrium points $_\pm K_\pm^0$:}
\begin{eqnarray}
  \bqz&=&\pm1 , \quad 
  \bqp=\bqz , \quad 
  \bc=0 ; \quad 
  \Omega_Y=0 ; \nonumber \\
  \nabla G &=&(0,0,-2\bqz,0) , \quad 
  (\bc \ {\rm eliminated}) ; \\
  \lambda_1 &=& 2\bqz , \quad 
  \lambda_2 =  3(2-\gamma)\bqz , \quad
  \lambda_3 = (3\gamma-2)\bqz . \nonumber
\end{eqnarray}
The point $_+K_+^0$ ($_-K^0_-$) is a local source (sink). Therefore there is 
a two-parameter set of orbits starting (ending) at the point $_+K_+^0$ 
($_-K^0_-$).

\paragraph{The equilibrium points $_\pm K_\mp^0$:}
\begin{eqnarray}
  \bqz&=&\pm1 , \quad 
  \bqp = -\bqz , \quad 
  \bc=0 ; \quad 
  \Omega_Y=0 ; \nonumber \\
  \nabla G &=&(0,0,2\bqz,0) , \quad 
  (\bc \ {\rm eliminated}) ; \\
  \lambda_1 &=& 2\bqz , \quad 
  \lambda_2 = -\lambda_3=(2-\gamma)\bqz . \nonumber
\end{eqnarray}
These two points are saddle points on the boundary and there are no asymptotes
into the physical phase space.

\subsubsection{Equilibrium points with intermediate tilt, $0<v^2<1$.}

\paragraph{The equilibrium points $_\pm\tilde{M}$:}
\begin{eqnarray}
  \bqz&=&\pm1 , \quad 
  \bqp=0 , \quad 
  \bc=1 , \nonumber \\
  v &=& \left[\frac{-\gamma(\gamma-1)+\sqrt{(\gamma-1)(\gamma^2(\gamma-1)+
  (2-\gamma)(3\gamma-2))}}{(3\gamma-2)(\gamma-1)}\right]\bqz ; \nonumber \\
  \Omega_Y&=&0 ; \nonumber \\
  \nabla G &=&\left( 0, -2\left[1+(\gamma-1)v^2\right], -2\gamma v, 0 \right)
  , \quad 
  (\bc \ {\rm eliminated} ) ; \\
  \lambda_1 &=& -2\bqz , \quad 
  \lambda_2 = \frac{2\bqz(\gamma-1)W}{(3\gamma-2)\left[1-(\gamma-1)v^2\right]} 
  , \nonumber \\
  \lambda_3 &=& \frac{(1+\bqz v)\left[(2-\gamma)+(3\gamma-2)\bqz v\right]}
  {-\gamma v} , \nonumber
\end{eqnarray}
where
\begin{equation}
  W = (3\gamma-2) + 4(3\gamma-4)v\bqz + (3\gamma^2-4)v^2 - (3\gamma-2)
  (\gamma-1)v^4 .
\end{equation}
These points correspond to the Minkowski spacetime. The constraint 
$\left|v\right|\leq1$ implies that these two points are physical only for 
$\gamma>\case65$. The point $_{+}\tilde{M}$ ($_{-}\tilde{M}$) enters the 
physical part of phase space though the point $_{+}M^+$ ($_{-}M^-$) 
(discussed below) at $\gamma=\case65$. Note the bifurcation in the eigenvalue 
for $\gamma=\case65$ in equation (\ref{eq:6/5}). As $\gamma$ increases, 
$_+\tilde{M}$ ($_-\tilde{M}$) moves along the line $_+M^+ - {_+M}^-$ 
($_-M^- - {_-M}^+$). From the expressions above, it follows that 
$\lambda_2 < 0$ and $\lambda_3 > 0 $ for $\case65 < \gamma < 2$. Therefore 
$_+\tilde{M}$ ($_-\tilde{M}$) is a saddle point in the full phase space. The 
positive (negative) eigenvalue $\lambda_3$ is associated with the vacuum 
submanifold so there is a two-dimensional separatrix surface ending (starting) 
at $_+\tilde{M}$ $(_-\tilde{M})$, the orbits of which lie in the interior 
phase space.

\subsubsection{Equilibrium points with extreme tilt, $v^2=1$.}

\paragraph{The equilibrium points $_\pm K_\pm^\pm$:}
\begin{eqnarray}
  \bqz&=&\pm 1 , \quad 
  \bqp=\bqz , \quad 
  \bc=0 ; \quad 
  \Omega_Y=0 ; \nonumber \\
  \nabla G &=& (0,-2\gamma \sgn(v), -2\gamma\bqz, 0) , \quad 
  (\bc \ {\rm eliminated}) ; \\
  \lambda_1 &=& 2\bqz , \quad 
  \lambda_2=4\bqz , \quad
  \lambda_3=-\frac{2(3\gamma-2)\bqz}{2-\gamma} . \nonumber
\end{eqnarray}
Note that the sign of $v$ affects the linearized constraint but not the 
stability of the points. These points are saddle points on the boundary and 
there are no asymptotes into the interior phase space.

\paragraph{The equilibrium points $_\pm M^\pm$:}
\begin{eqnarray}\label{eq:6/5}
  \bqz&=&\pm1 , \quad 
  \bqp=0 , \quad 
  \bc=1 , \quad 
  v=\bqz ; \quad 
  \Omega_Y=0 ; \nonumber \\
  \nabla G &=&(0,-2\gamma,-2\gamma\bqz,0) , \quad 
  (\bc \ {\rm eliminated} ) ; \\
  \lambda_1 &=& -2\bqz , \quad 
  \lambda_2 = -4\bqz , \quad
  \lambda_3 = \frac{-2(5\gamma-6)\bqz}{2-\gamma} . \nonumber
\end{eqnarray}
These points correspond to the Minkowski spacetime. Note that there is a 
bifurcation at $\gamma=\case65$. For $1< \gamma \leq \case65$ the points 
$_\pm M ^\pm$ are saddle points on the boundary and there are no asymptotes 
entering the interior of the phase space. At $\gamma=\case65$ they coincide 
with the points $_\pm\tilde{M}$ respectively. For $\case65 < \gamma < 2$ the 
point $_+ M ^+$ ($_- M^-$) is a sink (source) and there is a two-parameter 
set of orbits ending (starting) at the point.

\paragraph{The equilibrium points $_\pm\cal{H}^\mp$:}
\begin{eqnarray}
  \bqz&=&\pm1 , \quad 
  \bqp = -\bqz(1-\bc) , \quad 
  v=-\bqz ; \nonumber \\
  \Omega_Y &=& -2\bqz\bqp(1+\bqz\bqp) ; \nonumber \\
  \nabla G &=&(0,-2\gamma,2\bqz\gamma,-2(2-\gamma)\bqz\bqp(1+\bqz\bqp)) ; \\
  & & \qquad (\bc \ {\rm eliminated}) ; \nonumber \\
  \lambda_1 &=& 
  \lambda_2 = -2\bqz(1+2\bqz\bqp) , \quad
  \lambda_3 = 0 . \nonumber
\end{eqnarray}
The constraint $\Omega_Y\geq0$ implies that $\bqp$ is restricted to the 
interval $-1\leq\bqp\leq 0$ ($0\leq\bqp\leq 1$) when $\bqz=1$ ($\bqz=-1$).
There is only one non-zero eigenvalue correponding to two different 
eigenvectors. Equilibrium points on the line $_+\cal{H}^-$ ($_-\cal{H}^+$) 
are sources when $-1\leq\bqp<-\case12$ ($0\leq\bqp<\case12$) and sinks when 
$-\case12<\bqp\leq0$ ($\case12<\bqp\leq 1$).

\paragraph{The equilibrium points $_\pm K_\mp^\mp$:}
\begin{eqnarray}
  \bqz&=&\pm1 , \quad 
  \bqp=-\bqz , \quad 
  \bc=0 ; \quad 
  \Omega_Y=0 ; \nonumber \\
  \nabla G &=& (0,2\gamma \sgn(v), 2\gamma\bqz, 0) , \quad 
  (\bc \ {\rm eliminated}) ; \\
  \lambda_1 &=& \lambda_2 = 2 , \quad 
  \lambda_3 = 0 . \nonumber
\end{eqnarray}
These two points are the endpoints of the physical part of the lines 
$_{\pm}\cal{H}^\mp$, respectively. They describe the boundary asymptotes 
which play an important role in understanding the global dynamics of phase 
space. This motivates their special treatment. The points $_\pm K_\mp^\mp$ 
have no asymptotes entering the interior phase space.

\paragraph{The equilibrium points $_\pm M^\mp$:}
\begin{eqnarray}
  \bqz&=&\pm1 , \quad 
  \bqp = 0 , \quad 
  \bc=1 , \quad 
  v=-\bqz ; \quad 
  \Omega_Y=0 ; \nonumber \\
  \nabla G &=&(0,-2\gamma,2\gamma\bqz,0) , \quad 
  (\bc \ {\rm eliminated}) ; \\
  \lambda_1 &=& \lambda_2 = -2\bqz , \quad 
  \lambda_3 = 0 . \nonumber
\end{eqnarray}
These points correspond to the Minkowski spacetime, and are the other 
endpoints (compared to $_\pm K_\mp^\pm$) of the lines $_{\pm}\cal{H}^\mp$. 
These points have no asymptotes entering the interior phase space.

\subsection{Invariant submanifolds on the boundary of the SSS phase space}
\label{subsec:invsubmani}
The boundary is described by a number of invariant subsets: $\bqz=\pm1$, 
$\Omega_Y=1-\bqp^2-\bc^2=0$, $\bc=0$, $v=\pm1$. The submanifolds with 
$\bqz=\pm1$ are described by the same equations as the reduced equations for 
the plane symmetric type $_1$I models. The equations corresponding to the 
submanifolds with $v = \pm 1$ are identical to the reduced field equations 
for SSS spherically symmetric models with a neutrino fluid as the source. The 
submanifolds associated with $\Omega_Y=0$ are vacuum submanifolds with a 
test fluid. The constraint (\ref{eq:sssco}) implies that $\bqp\bc=0$ when 
$\Omega_Y=0$. If $\bqp=0$, one obtains the same equations as one has for 
SSS spherically symmetric vacuum models with a test perfect fluid, while 
$\bc=0$ leads to the same equations as those describing spatially 
homogeneous Kantowski-Sachs (KS) vacuum models with a test perfect fluid. 
Apart from the above vacuum submanifolds, $\bc=0$ also leads to a submanifold 
with $v=0$. This submanifold is described by the same equations as the 
orthogonal SH KS perfect-fluid models. The submanifolds are collected in 
Table 1 where we also introduce designations.

\begin{table}
  \caption{The various boundary submanifolds.}
  \begin{indented}
    \item[]
    \begin{tabular}{@{}cl}
      \br
      Boundary & Restriction \\
      \mr
      $_{{\pm}1}$I & $\bqz=\pm 1$ \\
      V & $\Omega_Y=0,\, \bqp=0,\, \bc = 1$ \\
      N$^\pm$ & $v = \pm 1$ \\
      KSV$_\pm$ & $\Omega_Y=0, \, \bqp=\pm 1, \, \bc=0$ \\
      KS & $\bc=0, v=0$ \\
      \br
    \end{tabular}
  \end{indented}
\end{table}

We will now describe the dynamical features of the individual boundary 
submanifolds.

\begin{figure}
  \centerline{\hbox{\epsfig{figure=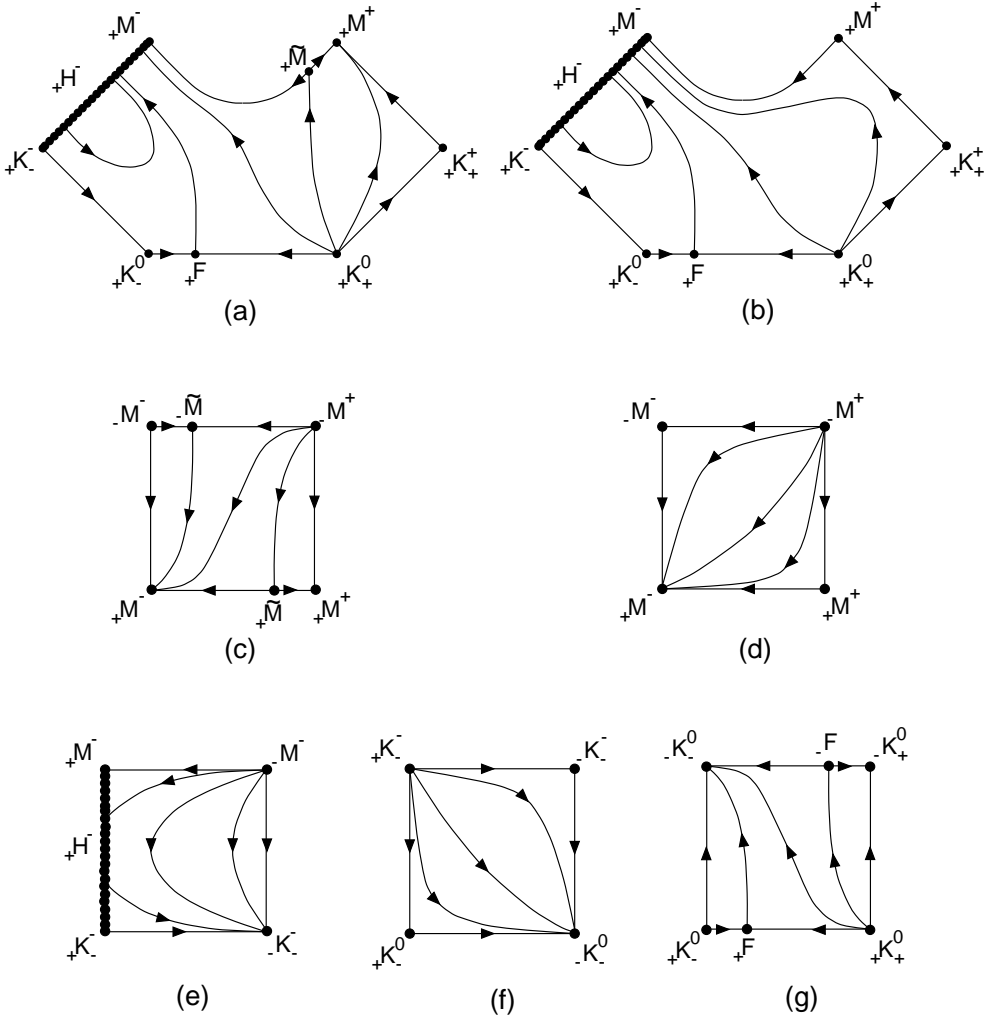, 
    width=1.0\textwidth}}}
  \caption{The phase portraits of some of the submanifolds that make up the 
    boundary of the reduced phase space. 
    (a) $_+{\rm I}$ for $\gamma > \case65$, 
    (b) $_+{\rm I}$ for $\gamma \leq \case65$, 
    (c) $V$ for $\gamma > \case65$, 
    (d) $V$ for $\gamma \leq \case65$, 
    (e) $N^-$, (f) $KSV_-$, (g) KS.}
\end{figure}

\paragraph{The $_{{\pm}1}$I submanifolds.}
The dynamical structure of the plane symmetric type $_{+1}$I submanifold is 
given in figure 1(a) ($\gamma > \case65$) and (b) ($\gamma \leq \case65$). 
That of $_{-1}$I is easily obtained by applying the discrete symmetry 
(\ref{eq:disc}).

\paragraph{The V submanifold.}
The dynamical structure of the V submanifold is given in figure 1(c) 
($\gamma > \case65$) and (d) ($\gamma \leq \case65$). Note that $\bqz$ is an 
increasing monotonic function.

\paragraph{The N$^\pm$ submanifolds.}
The dynamical structure of the N$^-$ submanifold is given in figure 1(e).
That of N$^+$ is easily obtained by applying the discrete symmetry  
(\ref{eq:disc}).
Note that $\bqp \leq 0$ ($\bqp \geq 0$) in the N$^-$ (N$^+$) case.
The N$^\pm$ submanifolds are solvable. The orbits are characterized by the 
following integral:
\begin{equation}
  \frac{\bqp\bc}{\left[\bqz-\bqp-\sgn(v)\bc\right]^2}={\rm constant} .
\end{equation}

\paragraph{The KSV$_\pm$ submanifolds.}
The dynamical structure of the KSV$_-$ submanifold is given in figure 1(f).
That of KSV$_+$ is easily obtained by applying the discrete symmetry
(\ref{eq:disc}). The variable $v$ satisfies the inequality $v\leq0$ ($v\geq0$)
in the KSV$_-$ (KSV$_+$) submanifold. The KSV$_\pm$ submanifolds are 
solvable. The integral describing the various orbits is given by
\begin{equation}
  \frac{v}{\left[1-\sgn(\bqp)\bqz \right]^\gamma}\left( 
  \frac{1-\bqz^2}{1-v^2} \right)^{\frac12(2-\gamma)} = {\rm constant} .
\end{equation}
It is worth pointing out that $\bqz$ is an increasing (decreasing) monotonic 
function and that $v$ is a decreasing (increasing) monotonic function for the 
KSV$_-$ (KSV$_+$) submanifold. 

\paragraph{The KS submanifold.}
Since the equations for this submanifold are the same as those for the SH KS 
models we automatically, as a bonus, get a dynamical description of these 
models. The SH KS models are of considerable interest and have been discussed 
in the literature before. Collins formulated the field equations as a 
dynamical system and made a qualitative analysis \cite{colKS}. However, he 
did not obtain a compact phase space. Uggla and von Zur M{\"u}hlen discussed 
the locally rotationally symmetric SH Bianchi type IX models \cite{uggmuh}. 
The KS models appear as a boundary submanifold in this context, but this was 
unfortunately not pointed out in \cite{uggmuh}. Neither was a phase portrait 
given. The dynamical structure of the KS submanifold is given in figure 1(g). 
Note that $\bqz$ is a monotonic function taking the solutions from big bang 
to big crunch. There exists a single time-symmetric orbit invariant under the 
discrete symmetry (\ref{eq:disc}). There is also a single orbit starting 
(ending) at the point $_+F$ ($_-F$). The $_+F$ point ($_-F$ point) 
corresponds to the flat FLRW solution in the SH KS geometry, and consequently 
these models have an asymptotically flat FLRW behaviour.

Orbits in the interior phase space coming close to the KS submanifold can be 
approximated with KS orbits.  The corresponding self-similar models in the 
interior phase space can thus be approximately described by a SH model 
multiplied by the homothetic factor, i.e. they are approximately conformally 
SH.

\section{Global dynamical behavior}
By appropriately `gluing together' the boundary submanifolds of the previous 
section, we obtain the reduced phase space shown in figure 2, of which the 
SSS models are orbits in the interior. By indicating the stability of the 
equilibrium points on the boundary we get figure 3 for $\gamma \leq \case65$ 
and figure 4 for $\gamma >\case65$. The orbits on the boundary corresponding 
to eigenvector directions of the points $_\pm F$ and $_\pm\tilde{M}$ are also 
shown. The two-dimensional separatrix surfaces entering the interior phase 
space from these two points will be discussed in section 4.3.

\begin{figure}
  \centerline{ \hbox{\epsfig{
    figure=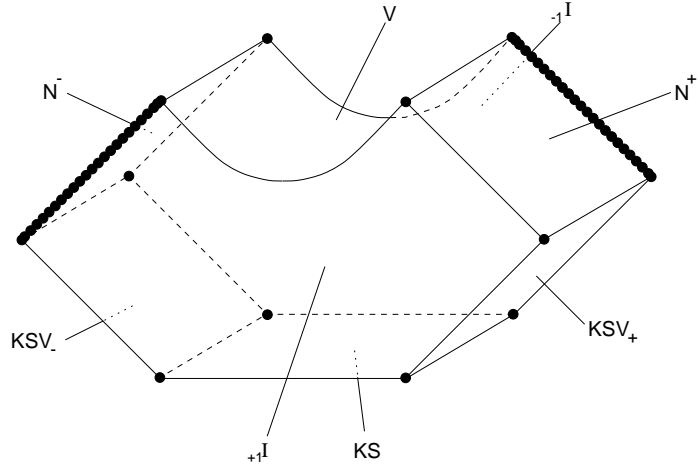, 
    width=0.7\textwidth}}}
  \caption{This picture shows how the boundary of the reduced phase space of 
    the spatially self-similar spherically symmetric models is constructed 
    out of the invariant submanifolds discussed in section 
    \ref{subsec:invsubmani}.}
\end{figure}

\begin{figure}
  \centerline{ \hbox{\epsfig{figure=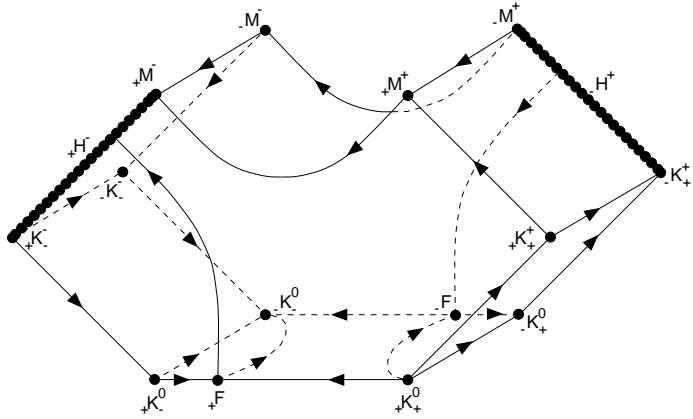, 
    width=0.7\textwidth}}}
  \caption{The reduced phase space for $\gamma\leq \case65$.}
\end{figure}

\begin{figure}
  \centerline{ \hbox{\epsfig{figure=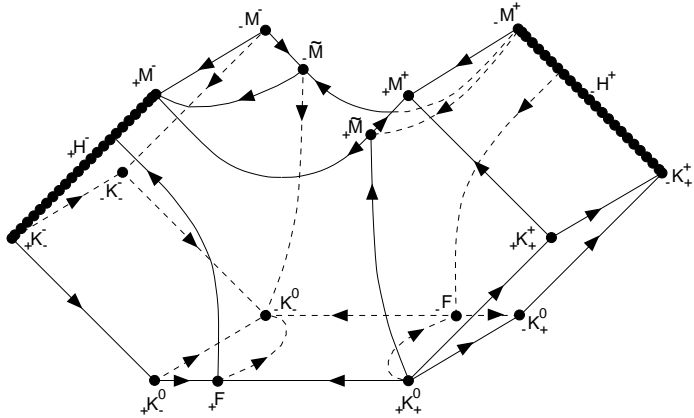, 
    width=0.7\textwidth}}}
  \caption{The reduced phase space for $\gamma> \case65$.}
\end{figure}

\subsection{Monotonic functions}
Monotonic functions are important tools for understanding the global 
dynamics. For example, the existence of a monotonic function in a region of 
phase space implies that there cannot be any limit cycles in that region. 

The function
\begin{equation}
  M = \left(1-\bqz^2\right)^{2-\gamma}\bqp\!^{3\gamma-4}\bc\!^{-\gamma}v^{2}
  \left( 1-v^2 \right)^{-(2-\gamma)} ,
\end{equation}
whose derivative is given by
\begin{equation}
M^\prime = -\left[\frac{(2-\gamma)(3\gamma-2)(1-v^2)\bc}{\gamma v}\right]M ,
\end{equation}
is monotone in the regions $v<0$ and $v>0$. There are no interior invariant 
submanifolds with $v=0$. Thus all attractors (in the present case, 
equilibrium points) lie on the boundary.

\subsection{Classification of orbits}
It is natural to divide the orbits in the SSS phase space into three 
categories depending on the extendibility properties into the timelike 
self-similar (TSS) region. In order to continue an orbit into the TSS region 
it must start or end at an equilibrium point with extreme tilt ($|v| = 1$). 
The classification presented here is based on whether orbits start and/or end 
at $_{\pm}{\cal H} ^\mp$. It is also possible to extend orbits into the TSS 
region through $_\pm M^\pm$, but from the structure of the TSS part of the 
phase space \cite{art:TSS} it is clear that these orbits are not as 
physically interesting as the orbits starting or ending at 
$_{\pm}{\cal H} ^\mp$. The three categories are:

0: Orbits that are purely SSS, i.e., orbits that do not start or end at 
$_{\pm}{\cal H}^{\mp}$.

I: Orbits that may be continued into the TSS region `once', e.g. orbits that 
start (end) at an equilibrium point not belonging to $_{\pm}{\cal H}^{\mp}$ 
while they end (start) at $_{\pm}{\cal H}^{\mp}$.

II: Orbits that may be continued into the TSS region `twice', e.g. orbits 
that start and end at $_{\pm}{\cal H}^{\mp}$.

\subsection{Invariant submanifolds in the interior of the SSS phase space}

\begin{figure}
  \centerline{ \hbox{\epsfig{figure=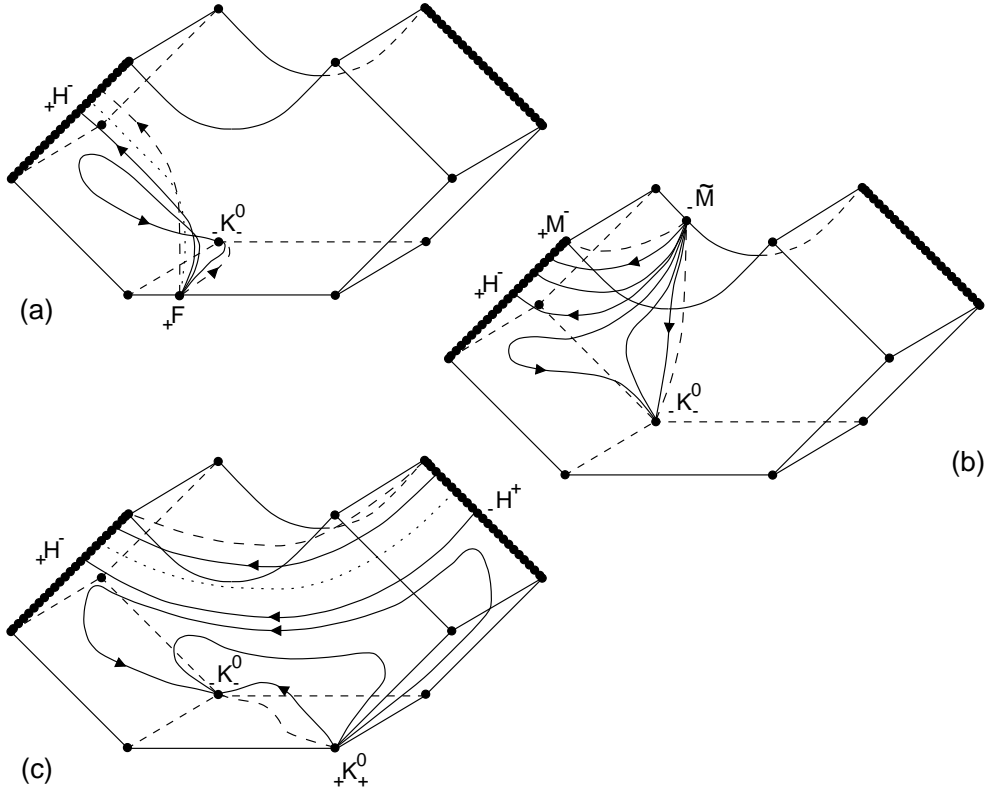, 
    width=1.0\textwidth}}}
  \caption{Separatrix submanifolds in the interior of the reduced phase space.
    a) The $_+F$ submanifold. The dotted orbit is the FLRW-orbit. 
    b) The $_-\tilde M$ submanifold. 
    c) The S submanifold. The dotted orbit is the TOVKMZ-orbit. 
    Dashed orbits lie in the boundary submanifolds.}
\end{figure}

{\it The $_{\pm}F$ submanifolds.} 
The separatrix surface spanned by the outgoing (ingoing) eigenvectors of the 
$_+F$ ($_-F$) point will be referred to as the $_+F$ ($_-F$) submanifold 
(figure 5(a)). The $_+F$ submanifold contains two different types of orbits. 
Near the $_{+1}$I submanifold, the orbits belong to category I and end 
in a band on the $_+{\cal H}^-$ line. The separatrix boundary in $_{+1}$I 
defines the upper boundary of the band. The lower boundary is for 
$\bqp = - {1 \over 2}$, where the stability of the $_+{\cal H}^-$ line 
changes to a source. The orbits in the $_+F$ submanifold that are close to 
the Kantowski-Sachs submanifold belong to category 0. They approach 
$_+{\cal H}^-$, but are repelled and end at $_-K^0_-$. The structure of the 
$_-F$ submanifold is easily obtained by applying the discrete symmetry 
(\ref{eq:disc}). Note that for the $_+F$ ($_-F$) separatrix, $v<0$ ($v>0$) in 
the whole interior.

In an SH slicing, an additional homothetic Killing vector leads to solutions 
being represented as equilibrium points \cite{janros}. This is the case with 
the flat FLRW solution. However, in the present case we have a slicing 
associated with the homothetic Killing vector, see figure 6. In this 
formulation, the flat FLRW solution appears as an one-dimensional orbit. The 
SSS part of the flat FLRW solution actually corresponds to two (equivalent) 
orbits contained within the `category I band'. The FLRW orbits can be 
found by imposing the Petrov type 0 condition along with vanishing fluid 
shear and acceleration. The orbits are characterized by
\begin{equation}
  \bqp = -\frac{2\bqz}{4+(3\gamma-2)v^2}, \quad C_1 = -\frac{3\gamma v\bqz}
{4+(3\gamma-2)v^2}.
\end{equation}
The FLRW orbit in the $_+F$ submanifold ends at a point on 
$_+\cal{H}^-$ with $\bqp = -2/(3\gamma+2)$. 

\begin{figure}
  \centerline{\hbox{\epsfig{figure=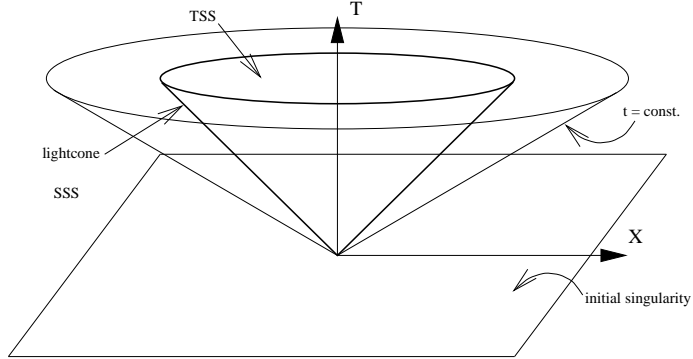, 
    width=0.7\textwidth}}}
  \caption{Schematic picture of the flat FLRW spacetime with a slicing 
    adapted to the homothetic Killing vector, compared to a slicing adapted 
    to the comoving time $T$. The cones are hypersurfaces of constant 
    self-similar variable $t$. The region inside the lightcone is the TSS 
    region. Note that the cones actually have a much more complicated 
    structure than in this over-simplified picture.}
\end{figure}

The invariant quantity $2m/R$ (see Appendix A.2) computed for the FLRW orbits
is given by
\begin{equation}
  \frac{2m}{R} = \frac13\tilde{\mu}R^2 = \frac{4}{(3\gamma-2)^2v^2} .
\end{equation}
Note that there is a one-parameter set of orbits within the $_+F$ submanifold 
lying arbitrarily close to the FLRW orbit throughout the entire SSS region.

Self-similar fluctuations in a flat FLRW universe have been studied by Carr 
and Yahil \cite{caryah}, although they focus their discussion on phenomena 
pertaining to the TSS region. In order to study density perturbations, we 
follow Ellis and Bruni \cite{ellbru}. Their approach has the advantage of 
being covariant and gauge independent. The fractional density gradient is 
characterized by a quantity $L$ (see \ref{sec:dens}). This quantity vanishes 
identically for non-tilted SH models. In the present case we have two such 
examples, namely the flat FLRW solution and the entire KS boundary 
submanifold. Solutions close to the flat FLRW in the $_+F$ submanifold can be 
parametrized by $L$: negative values of $L$ correspond to over-dense 
solutions ($\tilde{\mu}>\tilde{\mu}_{FLRW}$), and positive values correspond 
to under-dense solutions. Thus, the flat FLRW solution splits the $_+F$ 
submanifold into an under-dense part and an over-dense part. Over-dense 
solutions, starting out from the equilibrium point $_+F$ in a direction 
sufficiently close to the KS boundary, behave completely differently compared 
to the FLRW solution and correspond to category 0 orbits. The remaining 
orbits in the $_+F$ submanifold behave roughly as the flat FLRW orbit and 
belong to category I. Near the equilibrium point $_+F$, the fractional 
density gradient vanishes to linear order. Consequently, all solutions in the 
$_+F$ submanifold are near-homogeneous initially, with density fluctuations 
growing significant only at later stages, possibly with the exception of the 
origin $r=0$.

\paragraph{The $_{\pm}\tilde{M}$ submanifolds.}
The equilibrium points $_{\pm}\tilde{M}$ exist when $\gamma > \case65$. In 
this case one has additional separatrix surfaces in phase space. As for the 
$_+F$ submanifold, the orbits in the $_-\tilde M$ submanifold (figure 5(b)) 
are separated into two bands. Orbits starting close to the vacuum submanifold 
belong to category I and end on $_+{\cal H}^-$ in a continuous band from 
$_+M^-$ down to $\bqp = - {1 \over 2}$. The other band contains orbits 
belonging to category 0, which are repelled from $_+{\cal H}^-$ and end at 
$_-K^0_-$.

\paragraph{The Symmetric submanifold.} 
Orbits passing through $(\bqz=0,\bqp=0,\bc,v=0)$ constitute a submanifold 
containing `time-symmetric' solutions (figure 5(c)). The orbits in the 
symmetric submanifold (S submanifold) fall into two classes: those starting 
at $_-\cal{H}^+$ and ending at $_+\cal{H}^-$, thus belonging to category II, 
and those belonging to category 0, starting at $_+K_+^0$ and ending at 
$_-K_-^0$. 

One of the orbits in the S submanifold belonging to category II corresponds 
to the SSS part of a static self-similar solution associated with many names: 
Tolman \cite{tolman}, Oppenheimer and Volkoff \cite{oppie}, Klein 
\cite{klein} and Misner and Zapolsky \cite{misnerzapolsky}, see e.g. 
Henriksen and Patel \cite{patel}, Collins \cite{collins} and Bogoyavlensky 
\cite{bogoyavlensky}. We will call this orbit the TOVKMZ orbit. It is 
characterized by 
\begin{equation}
  \bqp = -\frac{2(\gamma-1)\bqz}{3\gamma-2} , \quad 
  \bc = -\frac{\gamma\bqz}{(3\gamma-2)v} .
\end{equation}
The quantity $2m/R$ takes the form
\begin{equation}
  \frac{2m}{R} = \tilde{\mu}R^2 = 
  \frac{4(\gamma-1)}{(2-\gamma)^2 + 8(\gamma-1)} .
\end{equation}

\subsection{General structure of interior orbits{\rm :}} 
Here we consider the behavior of orbits, not necessarily belonging to any of 
the submanifolds discussed above. A general orbit must start at one of the 
sources of the reduced phase space, and end at one of the sinks. The analysis 
is performed separately for $\gamma \leq \case65$ and $\gamma > \case65$. For 
$\gamma \leq \case65$, the possible sources are the point $_+K_+^0$, and the 
`source parts' of the lines $_+{\cal H}^-$ and $_-{\cal H}^+$. The only sinks 
are the point $_-K_-^0$, and the `sink parts' of the lines $_+{\cal H}^-$ and 
$_-{\cal H}^+$. Thus there are nine possible types of orbits for 
$\gamma \leq \case65$. It seems highly unlikely that orbits starting at 
$_+{\cal H}^-$ and ending at $_-{\cal H}^+$ exist, especially since the 
$_-K_-^0$ sink in general is very strong. Apart from such orbits, the 
existence of general orbits of all categories have been verified numerically. 
When $\gamma > \case65$, the appearance of the $_\pm {\tilde M}$ points 
changes the stability of $_-M^-$ to a source, and $_+M^+$ to a sink. Apart 
from all the types of orbits occuring for $\gamma \leq \case65$, there will 
be additional types of orbits. However, some of the new possible combinations 
of start- and end-points do not occur. This is due to the $_\pm {\tilde M}$ 
separatices `screening' the $_\pm M^\pm$ points. Thus, orbits starting at 
$_-M^-$ and ending at $_+M^+$ or $_-{\cal H}^+$ do not exist. Likewise, 
orbits starting at $_+{\cal H}^-$ and ending at $_+M^+$ do not appear either. 
Apart from these exclusions, the existence of general orbits of all 
categories have been verified numerically for $\gamma > \case65$ as well.

\section{Discussion}
In this paper we have used a dynamical systems approach to study the global 
structure of the solution space of SSS spherically symmetric perfect fluid 
models. By choosing variables adapted to the homothetic Killing vector field 
the phase space of these models can be made compact -- a very desirable 
property. As the constraint equation for these models is not solved globally, 
but rather locally at the equilibrium points of the system, we avoid 
`crushing' parts of phase space. As a result of having a complete regular 
phase space we find that there is a bifurcation at $\gamma=\frac{6}{5}$, 
something which has been overlooked in the previous literature. Also, a 
compact and regular phase space allows us to follow trajectories throughout 
their entire evolution. 

It is of interest to observe that the FLRW model appears as a trajectory in 
phase space. This is in stark contrast to Bianchi cosmology where the flat
FLRW appears as an equilibrium point. This affects the perturbation analysis
of this model. In Bianchi cosmology one only needs to know the phase space
in a neighbourhood of the equilibrium point. In the present case the FLRW
trajectory spans most of the phase space and thus there is a need to have a 
full phase space description to carry out a perturbation analysis. 

The existence of a monotonic function implies that there are no limit points 
in the interior of phase space. This stresses the importance of the boundary. 
The SH Kantowski-Sachs models appear as a boundary submanifold of the reduced 
phase space, and thus we obtain a compact description of these models as well.
  
We see that in the phase space corresponding to the SSS region there are 
models which are close to the FLRW orbit throughout their entire evolution in 
the SSS region. The models corresponding to these orbits can be interpreted 
as cosmological models with density perturbations. There are also orbits 
arbitrarily close to the static TOVKMZ orbit which also can be continued into 
the TSS region. To obtain a full spacetime picture of the models 
corresponding to these perturbed FLRW and static orbits, we will have to 
extend them into the TSS region. The TSS region will be studied in a 
forthcoming paper.

\appendix

\section{Interesting physical quantities}\label{sec:physquant}

\subsection{Fluid properties and Petrov type conditions}
The fluid expansion is given by
\begin{eqnarray}
  \tilde{\theta} &=& \frac{e^{-x}}{\sqrt{1-v^2}}\left\{\frac{v\dot{v}}{1-v^2}
  +\theta+3vB_1\right\} \nonumber \\
  &=& \frac{e^{-x}}{\sqrt{3}\sqrt{1-v^2}}\left\{\frac{\sqrt{3}v\dot{v}}
  {1-v^2} + 2\bar{\theta}+\bar{\sigma}_+ + 3\sqrt{3}vB_1\right\} \nonumber \\
  &=& \frac{e^{-x}}{\sqrt{3}(1-v^2)^{3/2}}\left\{vv^{\prime}+
  (1-v^2)(2\bqz+\bqp+3v\bc)\right\}Y .
\end{eqnarray}
The fluid shear is given by
\begin{eqnarray}
  \tilde{\sigma} &=& \frac{e^{-x}}{\sqrt{3}\sqrt{1-v^2}}\left\{\frac{v\dot{v}}
        {1-v^2} - \sigma_+\right\} \nonumber \\
  &=& \frac{e^{-x}}{3(1-v^2)^{3/2}}\left\{v\dot{v}
        - \bar{\theta}-2\bar{\sigma}_+\right\} \nonumber \\
  &=& \frac{e^{-x}}{3(1-v^2)^{3/2}}\left\{vv^{\prime}-(1-v^2)
        (\bqz+2\bqp)\right\}Y .
\end{eqnarray}
The fluid acceleration is given by
\begin{eqnarray}
  \tilde{a} &=& \frac{e^{-x}}{3\sqrt{1-v^2}}\left\{\frac{3\dot{v}}{1-v^2}
  +\theta v - 2\sigma_+v+3B_1\right\} \nonumber \\
  &=& \frac{e^{-x}}{3\sqrt{3}(1-v^2)^{3/2}}\left\{3\sqrt{3}v\dot{v}
  -3\bar{\sigma}_+v+3\sqrt{3}B_1\right\} \nonumber \\
  &=& \frac{e^{-x}}{\sqrt{3}(1-v^2)^{3/2}}\left\{v^{\prime}-(1-v^2)(\bqp v-\bc)
  \right\}Y .
\end{eqnarray}
The Weyl scalar is
\begin{eqnarray}
  C^2 &=& C_{abcd}C^{abcd} = \case49 e^{-4x} \left( 
  3{\dot \sigma }_+ - 3B_2{}^2 + \theta\sigma_+ - 2\sigma_+^2 \right)^2 
  \nonumber \\
  &=& \case49 e^{-4x} \left[ \sqrt{3}\left(\dot{\bt} + 
  2\dot{\bar \sigma}_+ \right) - 3B_2^2 - \bt\bsi - 2\bsi^2\right]^2 
  \nonumber \\
  &=& \case49 e^{-4x} \left[ \bqz^\prime + 2\bqp^\prime + F_Y\left( 
  \bqz + 2\bqp \right) \right. \nonumber \\
  & & \quad \left. + \bqz^2 - \bqp\left(\bqz + 2\bqp \right)-1\right]^2Y^4 .
\end{eqnarray}

Note that the magnetic part of the Weyl tensor is identically zero for all 
models. The spacetime is of Petrov type D if $C^2\neq0$ and Petrov type 0 
if $C^2=0$.

\subsection{The mass function}
For general spherically symmetric spacetimes the total mass-energy $m$ between
the centre distribution and some 2-space of symmetry is defined as, see e.g.,
Misner and Sharp \cite{misnersharpe}, Hernandez and Misner \cite{hernandez}
and Cahill and McVittie \cite{cahill},
\begin{equation}\label{eq:mr}
  \frac{2m}{R} =   1-
  \tilde{g}^{ab}\frac{\partial R}{\partial x^a}\frac{\partial R}{\partial x^b}
   = \frac{1+2\bqz\bqp + \bqp^2-\bc^2}{1-\bqz^2} ,
\end{equation}
where the parameter $R$ is the invariant radius of the symmetry surfaces.
By definition, the matter is outside the gravitational radius $2 m$ whenever
$2 m/R < 1$ and inside the gravitational radius when $2 m/R > 1$.
The latter case is associated with the existence of an apparent horizon
located at $2m/R=1$.

\subsection{Density perturbations}
\label{sec:dens}
A covariant approach to density perturbations has been given by Ellis and 
Bruni \cite{ellbru}. The fractional density gradient is defined as
\begin{equation}
  \Delta_a = \left({\Delta{\tilde \mu} \over {\tilde \mu}}\right)_a =
  {\tilde\mu}^{-1} {\tilde h}_a\!^{b} 
  {\tilde \partial}_b{\tilde \mu} .
\end{equation}
In the reduced phase space variables, this becomes
\begin{equation}
  \Delta_a = Y \, e^{-x} \, {L \over 1-v^2}\left(-v,1,0,0\right) ,
\end{equation}
where
\begin{equation}
  L = -{2\bc\left[1+(\gamma-1)v^2\right] + 
  \gamma v\left[2\bqz +(1+v^2)\bqp\right] \over 
  \sqrt{3}\left[1-(\gamma-1)v^2\right]} ,
\end{equation}
and $Y$ is the decoupled variable (\ref{eq:Y}).

\section{Coordinate and variable transformations}\label{sec:coord}
\setcounter{equation}{0}
Here we will give transformations to other coordinates and variables which 
have been used in the literature to study self-similar spherically symmetric 
models. The present variables lead to a compact and regular description 
everywhere in the SSS region. This is not the case with previously used 
variables, as is easily seen in the equations below. The simple algebraic 
relations make it easy to identify the points, or even manifolds, where the 
breakdowns occur.

\subsection{The comoving (fluid) approach}
The most commonly used approach is the fluid approach, where one uses a 
coordinate system adapted to the fluid velocity. The line element, of the 
presently considered SSS spherically symmetric models, takes the following 
form in the fluid approach (see, e.g., \cite{oripir}):
\begin{equation}
  d{\tilde s}^2 = - e^{\Psi(\lambda)}dT^2 + e^{\Lambda(\lambda)}dX^2 + 
  Y^2(\lambda)X^2 d\Omega^2 ,
\end{equation}
where $\lambda = X/T$ and $d\Omega^2 = d\theta^2 + \sin^2(\theta)d\varphi^2$.
The fluid velocity, $u^a$, is given by $e^{-\Psi/2}(1,0,0,0)$.

The following transformation:
\begin{equation}
  X = e^{x - F({\bar t})} , \qquad  T = e^{x-F(\bar t)+ \bar t} ,
\end{equation}
where
\begin{equation}
  \frac{dF}{d{\bar t}} = 
  -\frac{e^{\Psi + 2{\bar t}}}{e^\Lambda-e^{\Psi + 2{\bar t}}} ,
\end{equation}
leads to the `homothetic' adapted SSS spherically symmetric line element
\begin{equation}
  d{\tilde s}^2 = e^{2x} ds^2 = e^{2x}\left[
  -N^2d{\bar t}^2 + D_1{}^2 dx^2 + D_2{}^2 d\Omega^2\right] ,
\end{equation}
where
\begin{eqnarray}
  N^2 &=& e^{\Lambda + \Psi + 2{\bar t} -2F}
  (e^\Lambda-e^{\Psi + 2{\bar t}})^{-1} , \nonumber \\
  D_1^2 &=&
  e^{-2F}\left(e^\Lambda - e^{\Psi + 2{\bar t}}\right) , \quad
  D_2^2 = e^{-2F}Y^2 .
\end{eqnarray}
The fluid velocity $v$ is given by $v^2=e^{\Psi + 2{\bar t}-\Lambda}$.
Inserting this into the above expressions we find that
\begin{equation}
  \frac{dF}{d\bar{t}} = -\frac{v^2}{1-v^2}, \quad
  N^2=\frac{v^2e^{\Lambda-2F}}{1-v^2}, \quad
  D_1^2 = (1-v^2)e^{\Lambda-2F} .
\end{equation}
The fluid approach was used by Foglizzo and Henriksen \cite{foglizzo}, see
also Bicknell and Henriksen \cite{bicknell}, to write the self-similar 
spherically symmetric equations in a very simple way. Their variables are
related to those used in this paper according to
\begin{eqnarray}
  N_{F-H} &=& \frac{3\gamma^2v^2\mu_n}{4\left[ 1+
  \left(\gamma-1\right)v^2 \right]B_1^2} = \frac{3\gamma v\bsi}{2\sqrt{3}B_1}
         = \frac{3\gamma v\bqp}{2\bc} \\
  \mu_{F-H} &=& \frac{ \left[1+\left(\gamma-1\right)v^2\right]
  \left[\left(\bt+\bsi\right)^2 + 3B_2^2 - 3B_1^2 \right]}{\left(1-v^2\right)
  \mu_n} \nonumber \\
  &=& \frac{\sqrt{3}\gamma v\left[\left(\bt+\bsi\right)^2 + 3B_2^2 - 
  3B_1^2 \right]}{2B_1\bsi(1-v^2)} \nonumber \\ 
 &=&  \frac{3\gamma v\left(1+2\bqz\bqp +\bqp^2-\bc^2\right)}
 {2\bc\bqp(1-v^2)} \\
  v_{F-H} &=& v^{-1} .
\end{eqnarray}

\subsection{The Schwarzschild approach}
The usual way to represent a spherically symmetric line element is of the form
\begin{equation}\label{schwarz}
  ds^2 = -FdT^2 + GdR^2 + R^2d\Omega^2
\end{equation}
where, in the self-similar case, $F$ and $G$ are functions of $R/T$ only. The
following transformation:
\begin{equation}
  R = e^xB_2^{-1} , \quad T = e^{x+\phi(t)} ,
\end{equation}
where the function $\phi$ satisfies
\begin{equation}
  \frac{d\phi}{dt} = \frac{\sqrt{3}B_1^2}{\bar{\theta} +\bar{\sigma}_+} = 
  \frac{\bc^2 Y}{\sqrt{3}\left(\bqz+\bqp\right)} ,
\end{equation}
transforms the line element equation (\ref{schwarz}) into the homothetic 
form. The metric functions $F$ and $G$ can be written as
\begin{eqnarray}
  F &=& \frac{\left( \bar{\theta} + \bar{\sigma}_+\right)^2e^{-2\phi}}
  {B_1^2\left[3B_1^2 - \left(\bar{\theta} + \bar{\sigma}_+\right)^2\right]} = 
  \frac{3\left(\bqz+\bqp\right)^2 e^{-2\phi}Y^{-2}}
  {\bc^2\left[\bc^2 - \left(\bqz+\bqp\right)^2\right]} , \\
  G &=& \frac{3B_2^2}{3B_1^2 - \left(\bt + \bsi\right)^2}
  = \frac{1-\bqz^2}{\bc^2 - \left( \bqz + \bqp \right)^2} .
\end{eqnarray}
Identifying $2m/R = 1 - G^{-1}$ leads to equation (\ref{eq:mr}).

The radial 3-velocity $v_R$ of the fluid is given by
\begin{equation}
  v_R = \frac{\sqrt{3}vB_1 + \bt + \bsi}{\sqrt{3}B_1 + 
        v\left(\bt+\bsi\right)} = 
  \frac{v\bc + \bqz+\bqp}{\bc + v\left(\bqz + \bqp\right)} .
\end{equation}
In their treatment of self-similar spherically 
symmetric collapse, Ori and Piran \cite{oripir} used the following set of
variables:
\begin{eqnarray}
  \cal{M} &=& \frac{m}{R} = \frac{\left(\bt+\bsi\right)^2 + 3B_2^2 - 3B_1^2}
  {6B_2^2} = \frac{1+2\bqz\bqp + \bqp^2-\bc^2}{2\left(1-\bqz^2\right)} , \\
  D &=& \frac{\left(1-v^2\right)\mu_n}
  {2\left(1+(\gamma-1)v^2\right)B_2^2} = \frac{\sqrt{3}B_1\bsi(1-v^2)}
  {6\gamma v B_2^2} = \frac{\bc\bqp(1-v^2)}{2\gamma v(1-\bqz^2)} , \\
  u^r &=& \frac{\sqrt{3}vB_1 + \bt+\bsi}{\sqrt{3}B_2\sqrt{1-v^2}}
  = \frac{v\bc + \bqz+\bqp}{\sqrt{\left(1-\bqz^2\right)
  \left(1-v^2\right)}} .
\end{eqnarray}
The variables of Maison \cite{maison} are just
\begin{eqnarray}
  A_M^2 &=& \frac{3B_1^2 - \left(\bt+\bsi\right)^2}
  {3B_2^2} = 
  \frac{\bc^2 - \left(\bqz+\bqp\right)^2}{1-\bqz^2} , \\
  B_M^2 &=& \frac{3B_1^2}{\left(\bt+\bar{\sigma}_+\right)^2} = \frac{\bc^2}
  {\left(\bqz+\bqp\right)^2} , \\
  \tilde{\rho}_M &=& 4\pi R^2\tilde{\mu} = \frac{\left(1-v^2\right)\mu_n}
  {2B_2^2\left[1+(\gamma-1)v^2\right]} \nonumber \\
  & & = \frac{\left(1-v^2\right)\Omega_Y}
  {2\left(1-\bqz^2\right)\left[1+(\gamma-1)v^2\right]} , \\
  v_M &=& v_R .
\end{eqnarray}


\section*{References}

\begin{thebibliography}{00}

\bibitem{caryah}
Carr B J and Yahil A 1990 {\it Astrophys. J.} {\bf 360} 330

\bibitem{carcoley}
Carr B J and Coley A A Self-similarity in general relativity {\it Preprint}

\bibitem{cahtau71}
Cahill M E  and Taub A H 1971 {\it Commun. Math. Phys.} {\bf 21} 1

\bibitem{bogoyavlensky}
Bogoyavlensky O I 1985 {\it Methods in the Qualitative Theory of Dynamical
Systems in Astrophysics and Gas Dynamics} (Berlin: Springer)

\bibitem{carr}
Carr B J and Hawking S W 1974 {\it Mon.\ Not.\ R.\ Astron.\ Soc.} 
{\bf 168} 399

\bibitem{bicknell}
Bicknell G V and Henriksen R N 1978 {\it Astrophys. J.} {\bf 225} 237

\bibitem{oripir}
Ori A and Piran T 1990 {\it Phys. Rev. D} {\bf 42} 1068

\bibitem{koike}
Koike T, Hara T and Adachi S 1995 {\it Phys. Rev. Lett.} {\bf 74} 5170

\bibitem{maison}
Maison D 1996 {\it Phys. Lett.} {\bf 366B} 82

\bibitem{coley}
Coley A A 1997 {\it Class. Quantum Grav.} {\bf 14} 87

\bibitem{moschetti}
Moschetti G 1987 {\it Gen.\ Rel.\ Grav.} {\bf 19} 155

\bibitem{anile}
Anile A M, Moschetti G and Bogoyavlenski O I 1987 {\it J.\ Math.\ Phys.} 
{\bf 28} 2942

\bibitem{patel}
Henriksen R N and Patel K 1991 {\it Gen.\ Rel.\ Grav.} {\bf 23} 527

\bibitem{ear}
Eardley D M 1974 {\it Commun. Math. Phys.} {\bf 37} 287

\bibitem{wu}
Wu C 1981 {\it Gen.\ Rel.\ Grav.} {\bf 13} 625

\bibitem{hhss}
Nilsson U S and Uggla C 1997 {\it Class. Quantum Grav.} {\bf 14} 1965

\bibitem{art:TSS}
Goliath M, Nilsson U S and Uggla C 1997 in preparation

\bibitem{kramer}
Kramer D, Stephani H, MacCallum M and Herlt E 1980 
{\it Exact Solutions of Einstein's Field Equations} 
(Cambridge: Cambridge University Press)

\bibitem{ssslrs}
Nilsson U and Uggla C 1996 {\it Class. Quantum Grav.} {\bf 13} 1601

\bibitem{hewitt}
Hewitt C G and Wainwright J 1992 {\it Phys. Rev. D} {\bf 46} 4242

\bibitem{colKS}
Collins C B 1977 {\it J.\ Math.\ Phys.} {\bf 18} 2116

\bibitem{uggmuh}
Uggla C and von Zur M\"uhlen H 1990 {\it Class. Quantum Grav.} {\bf 7} 1365

\bibitem{janros}
Jantzen R T and Rosquist K 1986 {\it Class. Quantum Grav.} {\bf 3} 281

\bibitem{ellbru}
Ellis G F R and Bruni M 1989 {\it Phys. Rev. D} {\bf 40} 1804

\bibitem{tolman}
Tolman R C 1939 {\it Phys. Rev.} {\bf 55} 364

\bibitem{oppie}
Oppenheimer J R and Volkoff G M 1939 {\it Phys. Rev.} {\bf 55} 374

\bibitem{klein}
Klein O 1947 {\it Ark. Mat. Astr. Fys.} {\bf 34A} N:o 19 1

\bibitem{misnerzapolsky}
Misner C W and Zapolsky H S 1964 {\it Phys. Rev. Lett.} {\bf 12} 635

\bibitem{collins} 
Collins C B 1985 {\it J.\ Math.\ Phys.} {\bf 26} 2268

\bibitem{misnersharpe}
Misner C W and Sharp D H 1964 {\it Phys. Rev. B} {\bf 136} 571  

\bibitem{hernandez}
Hernandez W C and C W Misner 1966 {\it Astrophys. J.} {\bf 143} 452

\bibitem{cahill}
Cahill M E and G C McVittie 1970 {\it J.\ Math.\ Phys.} {\bf 11} 1382

\bibitem{foglizzo}
Foglizzo T and Henriksen R N 1993 {\it Phys. Rev. D} {\bf 48} 4645

\end{thebibliography}
\end{document}